# Solar Analogs as a Tool to Understand the Sun

Allison Youngblood, Steve Cranmer, Sam Van Kooten, James Paul Mason, J. Sebastian Pineda, Kevin France, Dmitry Vorobiev, Frank Eparvier, Yuta Notsu

**Summary:** Solar analogs, broadly defined as stars similar to the Sun in mass or spectral type, provide a useful laboratory for exploring the range of Sun-like behaviors and exploring the physical mechanisms underlying some of the Sun's most elusive processes like coronal heating and the dynamo. We describe a series of heliophysics-motivated, but astrophysics-like studies of solar analogs. We argue for a range of stellar observations, including (a) the identification and fundamental parameter determination of new solar analogs, and (b) characterizing emergent properties like activity, magnetism, and granulation. These parameters should be considered in the framework of statistical studies of the dependences of these observables on fundamental stellar parameters like mass, metallicity, and rotation.

**Motivation:** Many of the long-standing problems in solar physics are difficult to answer because we only have a single instance of the Sun. The Sun is one of the few objects in astronomy studied as the only representative of its class. We should strive to understand the Sun based on ensemble average properties as is done for galaxies, planets, interstellar clouds, and more. Solar analogs, or stars that are solar-like in mass and spectral type, can allow us to compare the Sun to the ensemble average properties of solar analogs and probe the range of possible Sun-like behaviors *by effectively varying one or more fundamental variables at a time*. For example, the physical processes that govern the dynamo and coronal heating will be better understood if we could empirically determine how they depend on factors such as the Sun's mass and rotation rate, which determine the size of its convective zone and basal coronal field complexity, respectively. By studying other stars, we may determine how these processes depend on effective temperature, metallicity, and age. Applications of solar analog studies extend beyond heliophysics to planetary science and astrophysics, particularly for understanding (exo)planet atmospheres and habitability, the influence of stars on the interstellar medium, and comparable plasma environments such as jets and galactic winds.

**Current state:** Solar analogs (e.g., G type stars; 0.9-1.1 solar masses; ~5400-5900 K), have long been recognized for their utility for solar physics (e.g., the Mount Wilson HK survey; Wilson 1978). Some examples of recent solar analog studies include the Ribas et al. (2005) Sun-in-Time study, which examined six G1-G5 stars of a variety of ages in order to examine the Sun's evolution of X-ray and UV activity with age. However, stellar models indicate there may be multiple possible evolutionary tracks for solar analogs (Tu et al. 2015). Examining Sun-like stars with *Kepler*, astronomers have also constrained the Sun's superflare rate (Maehara et al. 2012), found that the Sun's dynamo may be in a state of transition (Metcalfe & van Saders 2016), and shown that the surface manifestations of the Sun's convection and dynamo appear to be less active than for most Sun-like stars (e.g., Bastien et al. 2013; Reinhold et al. 2020).

**From now to 2050:** By 2050, we can create a large database of solar analogs ranked based on their similarity to the Sun with many measured properties, and perform statistical analyses and modeling to determine how higher-level or emergent properties (like activity) of the solar analogs depend on fundamental properties like mass. Obtaining so many fundamental and emergent properties for hundreds and possibly thousands of stars is truly a multi-decade effort and will rely on next generation telescopes for some of the more distant targets and difficult observables.

**Phase A: Identifying a large sample of solar analogs**
Building on the already large sample of nearby solar analogs and refining their parameters is possible with *Gaia*. Fundamental parameters like mass, radius, effective temperature, spectral type, metallicity, age, rotation, multiplicity, and planet architectures will need to be uniformly determined to identify the best targets. Many of these parameters are provided by *Gaia*'s second data release (DR2) and upcoming *Gaia* data releases, but the best solar analog targets (e.g., G type stars of any age) will require more detailed follow up and verification. This follow-up and verification effort will require many observations and may take 5-10 years.

**Phase B: Measure parameters of interest for the solar analog sample**
Characterizing many stars for emergent properties will be necessary to discover how the Sun's physical processes depend on fundamental parameters (Phase C). Fortunately some of the data required for these parameters will already have been obtained in Phase A. For example, global magnetic field strength via Zeeman broadening requires high resolution optical spectroscopy. However, additional multi-wavelength observational capabilities will be required to characterize the full complement of emergent features, including long-term polarimetric, spectroscopic, and photometric monitoring of individual stars. We summarize the relevant observations in Table 1.

As there are numerous emergent properties that are of interest to the community, here we highlight two possible solar analog studies that exemplify the required work in the coming decades:

1) *Precise, long-term optical photometric monitoring:* this dataset (in the vein of *Kepler, TESS,* and ESA's upcoming PLATO mission) would enable a precise determination of fundamental parameters like mass, radius, rotation period, and age, and would also provide emergent properties like flicker (granulation), spot/faculae coverage, and constraints on differential rotation and the butterfly diagram (e.g., Silva-Valio &

Lanza 2011; Bazot et al. 2018)

2) *X-ray, EUV, and FUV spectrophotometry:* this dataset would provide coronal and chromospheric heating rates and constrain theory on the underlying processes, as well as provide information useful for assessing the impacts on exoplanet atmospheres and the atmospheric histories of solar system planets like Venus and Mars.

*Table 1 - summary of observations needed and stellar parameters retrievable. Most of these parameters require long-term monitoring.*

| Observation Requirement | Stellar parameters |
|---|---|
| High-resolution optical spectroscopy | Effective temperature, spectral type, metallicity<br>Magnetic activity, magnetic field strength and/or topology (polarimetry required), activity cycles<br>Planetary companions and multiplicity<br>Jitter / Granulation |
| High-precision optical photometry | Flicker / Granulation<br>Mass, radius, rotation period, age |
| X-ray / UV spectrophotometry | Radiation environment<br>Activity / activity cycles<br>Coronal heating |

**Phase C: Determine emergent properties' correlation with fundamental parameters**
Using statistical techniques like principal component analysis, individual studies can explore how changes in the Sun's mass, temperature, metallicity, rotation, age, multiplicity etc. affect observables like activity, luminosity, and magnetism/dynamo. For example, we could directly explore what happens to coronal activity if you increase the rotation rate by 50%. By comparing observations of a statistical sample of solar analogs to model predictions, we could fine tune our understanding of the solar dynamo. Overall, the large, uniform database created in Phase A will enable a wide range of studies by a wide range of groups within heliophysics as well as within the other NASA science divisions.